**Comment on "Cosmic Bell Test: Measurement Settings from Milky Way Stars"**

Ref. [1] reports on observed violations of the Bell-CHSH inequality regarding correlations between the measured polarizations of distant photons, with the independence of the settings of the polarizers ensured by choosing them according to the frequencies of starlight photons. This remarkable achievement demonstrates once more how quantum phenomena force us to cast doubt on some of our most firmly held metaphysical suppositions, which are indeed discussed at length in [1]. Unfortunately, two statements are presented as basic assumptions of the analysis, whereas each follows from more basic assumptions. These apparent misrepresentations do not originate in Ref. [1] – in fact, they dominate the recent literature. It is thus crucial to clarify the issues.

One such statement is "the assumption that there are no statistical correlations between the choices of measurement settings and anything else that can causally affect the measurement outcomes." This is not a fundamental assumption of the derivation of Bell's theorem. It follows from: (i) it is appropriate to use free variables to describe the measurement settings (this is associated with "free will," see Ref. 22 of [1]); and (ii) the causal arrow of time. In mathematical language, the lack of statistical correlations between the measurement settings, denoted $a$ and $b$, and the previous state of the system, described by $\lambda$, is due to *mutual independence*: (i) $a$ and $b$ are independent of $\lambda$ because they are free variables; (ii) $\lambda$ is independent of $a$ and $b$ because it describes *the past* with respect to them.

The second statement is "objects possess complete sets of properties on their own, prior to measurement." This is called "realism" in [1], but as this word has been used with different meanings in the present context [2], it is better to label it as "determinism:" the properties $\lambda$, together with the settings $a$ and $b$, determine the measurement results $A$ and $B$, not merely their probabilities. Bell himself expressed frustration at the difficulty in getting across the point that "to the limited degree to which <u>determinism</u> plays a role in the EPR argument, it is not assumed but <u>inferred</u>" [3] (emphases in original). The inference, part of the EPR argument, follows from (i) the assumption of local causality; and (ii) the perfect correlations predicted by Quantum Mechanics (QM) for certain pairs of particles. Of course, determinism *was* simply assumed at times, e.g., in [4], but it was soon shown that the same result holds for indeterminism as well [5]. The analysis also applies to models with no hidden variables ($\lambda$ is then a constant). Thus, in a framework which presupposes the causal arrow of time and the use of free variables for measurement settings, no local model can reproduce the predictions of QM, or agree with observations.

Which of the assumptions should be rescinded? Pertinent toy models which violate either causality [6],[7] or measurement independence [8] have been suggested. The latter was presented as if violations of measurement independence are equivalent to violations of "free will" (as causality was taken for granted), and was quoted prominently in [1]. However, Eq. (8) of Ref. [8] is explicitly retrocausal – the distribution of $\lambda$ depends directly on the settings – and the model ostensibly violates "free will" only as a result of additional steps, which may be questioned. While this may be controversial, we should at least strive for consensus on the much simpler matter of what the basic assumptions of the analysis are. It is often stated that Bell's theorem reveals a tension between QM and relativity, or between QM, relativity and free will, but it is relativistic causality which is involved, rather than the time-reversal symmetric aspects of relativity. Realism or determinism is not a necessary assumption in deriving the theorem, but the causal arrow of time is.

Discussions with K. Wharton, M.J.W. Hall and A.S. Friedman are gratefully acknowledged.

N. Argaman*

Physics Department,
Nuclear Research Center – Negev
P.O. Box 9001 Be'er Sheva 84190, Israel

Version of May 11, 2017.